\documentclass[12pt,a4paper]{article}

\usepackage{epsfig}
\usepackage{amsmath}
\usepackage{amssymb}
\usepackage{verbatim}
\usepackage{bm}
\usepackage{dsfont}
\usepackage[footnotesize]{caption}
\usepackage{subfigure}
\usepackage{cancel}
\usepackage{cite}

\begin{document}

\begin{flushleft}
DESY 12-032\\
March 2012
\end{flushleft}

\vskip 1cm

\begin{center}
{\LARGE\bf WIMP Dark Matter from\\[2mm] 
Gravitino Decays and Leptogenesis}

\vskip 2cm

{\large W.~Buchm\"uller, V.~Domcke,  K.~Schmitz}\\[3mm]
{\it{
Deutsches Elektronen-Synchrotron DESY, 22607 Hamburg, Germany}
}
\end{center}

\vskip 1cm

\begin{abstract}
\noindent 
The spontaneous breaking of $B$$-$$L$ symmetry naturally accounts for the
small observed neutrino masses via the seesaw mechanism. We have recently shown that the
cosmological realization of $B$$-$$L$ breaking in a supersymmetric theory 
can successfully generate the initial conditions of the hot early universe, 
i.e.\ entropy, baryon asymmetry and dark matter, if the gravitino is the 
lightest superparticle (LSP). This implies relations between neutrino
and superparticle masses.
Here we extend our analysis to the case of very heavy gravitinos which
are motivated by hints for the Higgs boson at the LHC. We find that
the nonthermal production of `pure' wino or higgsino LSPs, i.e.\ weakly interacting
massive particles (WIMPs), in heavy gravitino decays 
can account for the observed amount of dark matter
while simultaneously fulfilling the constraints imposed by primordial nucleosynthesis
and leptogenesis
within a range of LSP, gravitino and neutrino masses. For instance, a mass of the lightest neutrino
of $0.05~\mathrm{eV}$ would require a higgsino mass below $900~\mathrm{GeV}$
and a gravitino mass of at least $10~\mathrm{TeV}$.
\end{abstract}.

\thispagestyle{empty}

\newpage

\subsubsection*{Introduction}

We have recently proposed that the spontaneous breaking of $B$$-$$L$, the
difference of baryon and lepton number, sets the initial conditions of the
hot early universe \cite{Buchmuller:2010yy,Buchmuller:2012wn}. In a
supersymmetric extension of the Standard Model, with $B$$-$$L$ breaking
at the grand unification (GUT) scale, an initial phase of unbroken $B$$-$$L$
yields hybrid inflation, ending in tachyonic preheating during which
$B$$-$$L$ is spontaneously broken. If the gravitino is the lightest 
superparticle (LSP), entropy, baryon asymmetry and gravitino dark matter can be 
produced in the subsequent reheating process. Successful baryogenesis via
leptogenesis and the generation of the observed relic dark matter density
require relations between neutrino masses and superparticle masses, in
particular a lower bound of $10~\mathrm{GeV}$ on the gravitino mass
\cite{Buchmuller:2012wn}.

In this Letter we want to point out that the spontaneous breaking of
$B$$-$$L$ can also ignite the thermal phase of the universe if the gravitino
is the heaviest superparticle. This possibility is realized in anomaly
mediation \cite{Giudice:1998xp,Randall:1998uk} and has recently been
reconsidered in the case of wino \cite{Ibe:2011aa}, higgsino 
\cite{Jeong:2011sg} and bino \cite{Krippendorf:2012ir} LSP, motivated
by hints of the LHC experiments ATLAS and CMS that the Higgs boson
may have a mass of about $125~\mathrm{GeV}$ 
\cite{Aad:2012si,Chatrchyan:2012tx}. It is known that a gravitino heavier 
than about $10~\mathrm{TeV}$ can be consistent with primordial nucleosynthesis
and leptogenesis \cite{Weinberg:1982zq,Gherghetta:1999sw,Ibe:2004tg}.
In the following we shall discuss the restrictions on the mass of a weakly interacting massive
particle (WIMP) as LSP, which
are imposed by the consistency of hybrid inflation, leptogenesis, big bang nucleosynthesis (BBN) and
the dark matter density.

\subsubsection*{Spontaneous $B$$-$$L$ breaking as the origin of the hot early universe}

Our starting point is the supersymmetric standard model with right-handed 
neutrinos and spontaneous $B$$-$$L$ breaking, described by the superpotential
\begin{equation}
\label{eq_W}
 W = \frac{\sqrt{\lambda}}{2} \, \Phi \, (v_{B-L}^2 - 2 \, S_1 S_2) + 
\frac{1}{\sqrt{2}} h_i^n n_i^c n_i^c S_1 + 
h^{\nu}_{ij} \textbf{5}^*_i n_j^c H_u + W_{\text{MSSM}} \,.
\end{equation}
Here $S_1$ and $S_2$ are the chiral superfields containing the Higgs superfield $S$ 
which breaks $B$$-$$L$ at the scale $v_{B-L}$, $\Phi$ contains the inflaton, 
i.e.\ the scalar field driving inflation, and $n_i^c$ denote the superfields 
containing the charge conjugates of the right-handed neutrinos; 
$h$ and $\lambda$ are coupling constants, and
$W_{\text{MSSM}}$ is the superpotential of the minimal supersymmetric standard
model with quarks, leptons and Higgs fields.
The requirement of consistency with hybrid inflation fixes the scale of $B$$-$$L$
breaking to a value close to the GUT scale, $v_{B-L} = 5 \times 10^{15}$~GeV, cf.\ Ref.~\cite{Buchmuller:2012wn}.
The superfields are arranged
in $SU(5)$ multiplets, i.e.\ $\textbf{5}_i^* = (d^c_i, \, l_i)$, $i = 1,2,3$, 
and we assume that the colour triplet partners of the electroweak Higgs doublets
$H_u$ and $H_d$ have been projected out. The vacuum expectation values 
$v_u = \langle H_u \rangle$ and $v_d = \langle H_d \rangle$ break the 
electroweak symmetry. In the following we will assume large 
$\tan \beta = v_u/v_d$, implying 
$v_d \ll v_u \simeq v_{EW} = \sqrt{v_u^2 + v_d^2}$.

\begin{figure}
\begin{center}
\includegraphics[width=12cm]{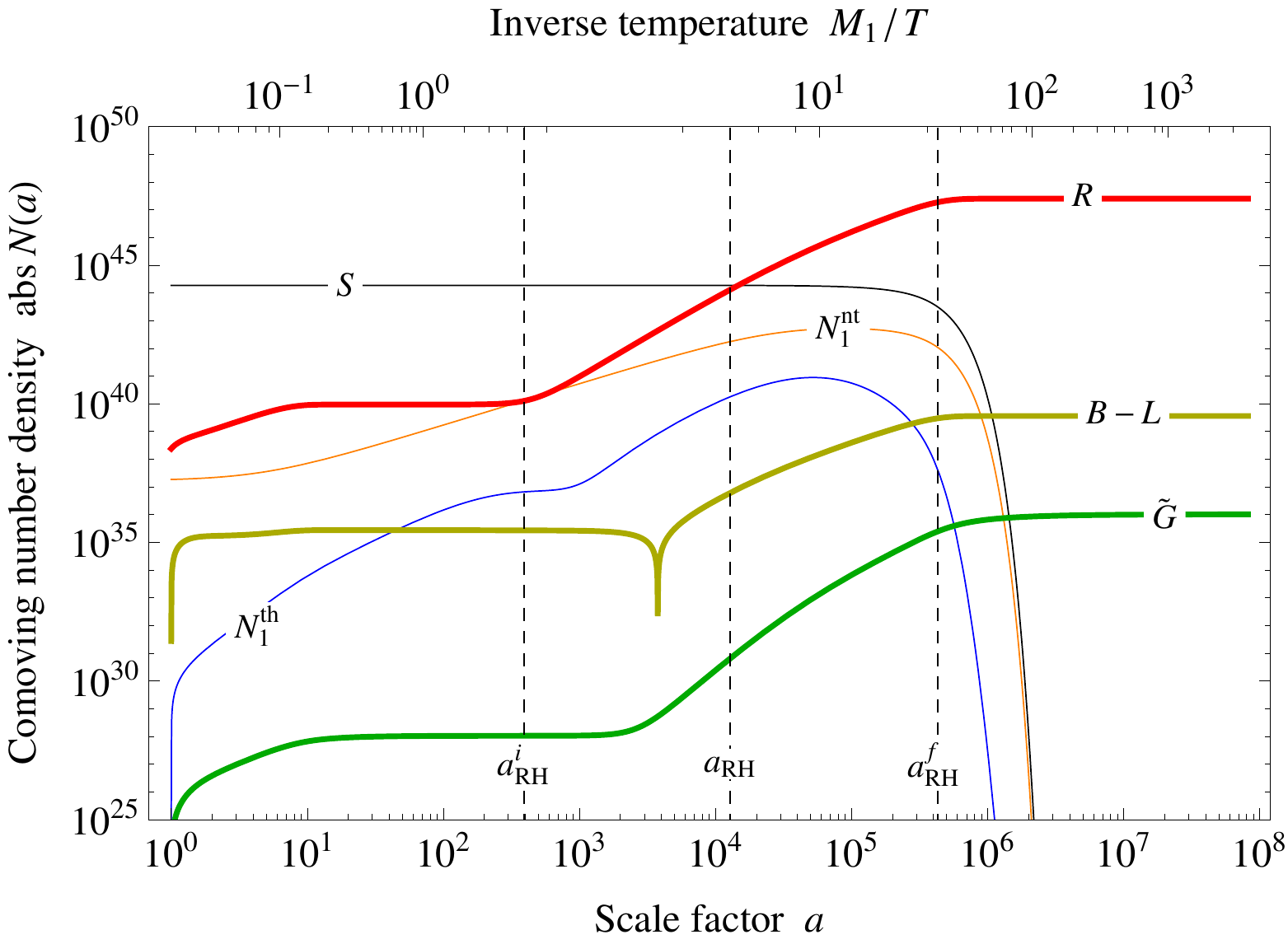}

\vspace*{5mm}

\includegraphics[width=12cm]{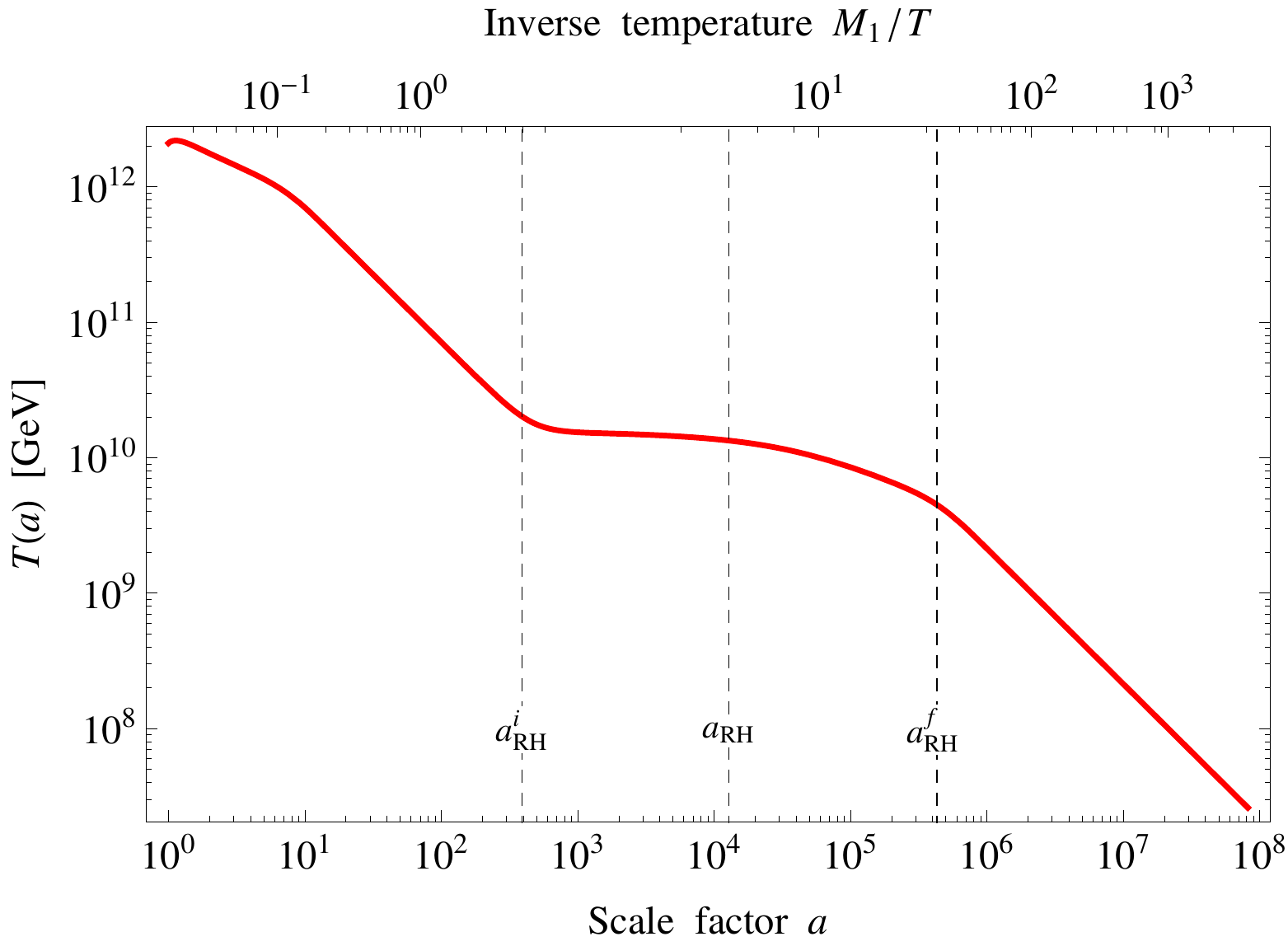}
\caption{{\bf Upper panel:} Comoving number densities of Higgs bosons ($S$), 
thermally and nonthermally produced heavy neutrinos ($N_1^{\mathrm{th}}$, 
$N_1^{\mathrm{nt}}$), radiation ($R$), lepton asymmetry ($B$$-$$L$) and gravitinos ($\widetilde G$). 
{\bf Lower panel:} Emergent plateau of approximately constant reheating
temperature.
Input parameters: Heavy neutrino mass $M_1 = 1\times 10^{11}~\mathrm{GeV}$,
effective neutrino mass $\widetilde{m}_1 = 4\times 10^{-2}~\mathrm{eV}$.
The $B$$-$$L$ scale is fixed by requiring consistency with hybrid inflation,
$v_{B-L} = 5\times 10^{15}~\mathrm{GeV}$.}
\label{fig:reheat}
\end{center}
\end{figure}

The Yukawa couplings are conveniently parametrized in terms of 
Froggatt-Nielsen flavour charges, cf.\ Ref.~\cite{Buchmuller:2012wn}, which govern the
hierarchy of quark and lepton masses and mixings.
For simplicity, we restrict our analysis to the case of hierarchical 
heavy neutrino masses $M_i$ and a heavy Higgs boson multiplet $S$, 
$m_S = M_3 = M_2 = M_1/\eta^2$, where
$\eta \simeq 1/\sqrt{300}$ is the hierarchy parameter of the Froggatt-Nielsen
flavour model.

The most important parameters for the
reheating process are the masses and vacuum decay widths of $S$ and $N_1$, which
can be expressed in terms of $M_1$ and the effective neutrino mass
$\widetilde{m}_1$,
\begin{equation}
\Gamma_S^0 = \frac{1}{32 \pi} \frac{M_1^2}{v_{B-L}^2} \ m_S
\left( 1 - 4 \frac{M_{1}^2}{m_S^2} \right)^{1/2}, \quad
\Gamma^0_{N_{1}} = \frac{1}{4\pi} \left(h^{\nu\,\dagger}h^\nu\right)_{11} M_1
= \frac{1}{4\pi}\ \frac{\widetilde{m}_1 M_1}{v_{EW}^2} \,M_1\ .
\end{equation}
Varying $M_1$ corresponds to varying one of the flavour charges. The 
uncertainty in $\widetilde{m}_1$ is related to unknown $\mathcal{O}(1)$
coefficients in the Froggatt-Nielsen model; a typical value is
$\widetilde{m}_1 \sim 0.04~\mathrm{eV}$ \cite{Buchmuller:2011tm}.
It is well known that $\widetilde{m}_1$ is bounded from below by the
lightest neutrino mass $m_1$ \cite{Fujii:2002jw}. Consequently, constraints
on $\widetilde{m}_1$ directly translate into constraints on the light neutrino
mass spectrum.

\begin{figure}
\begin{center}
\includegraphics[width=10cm]{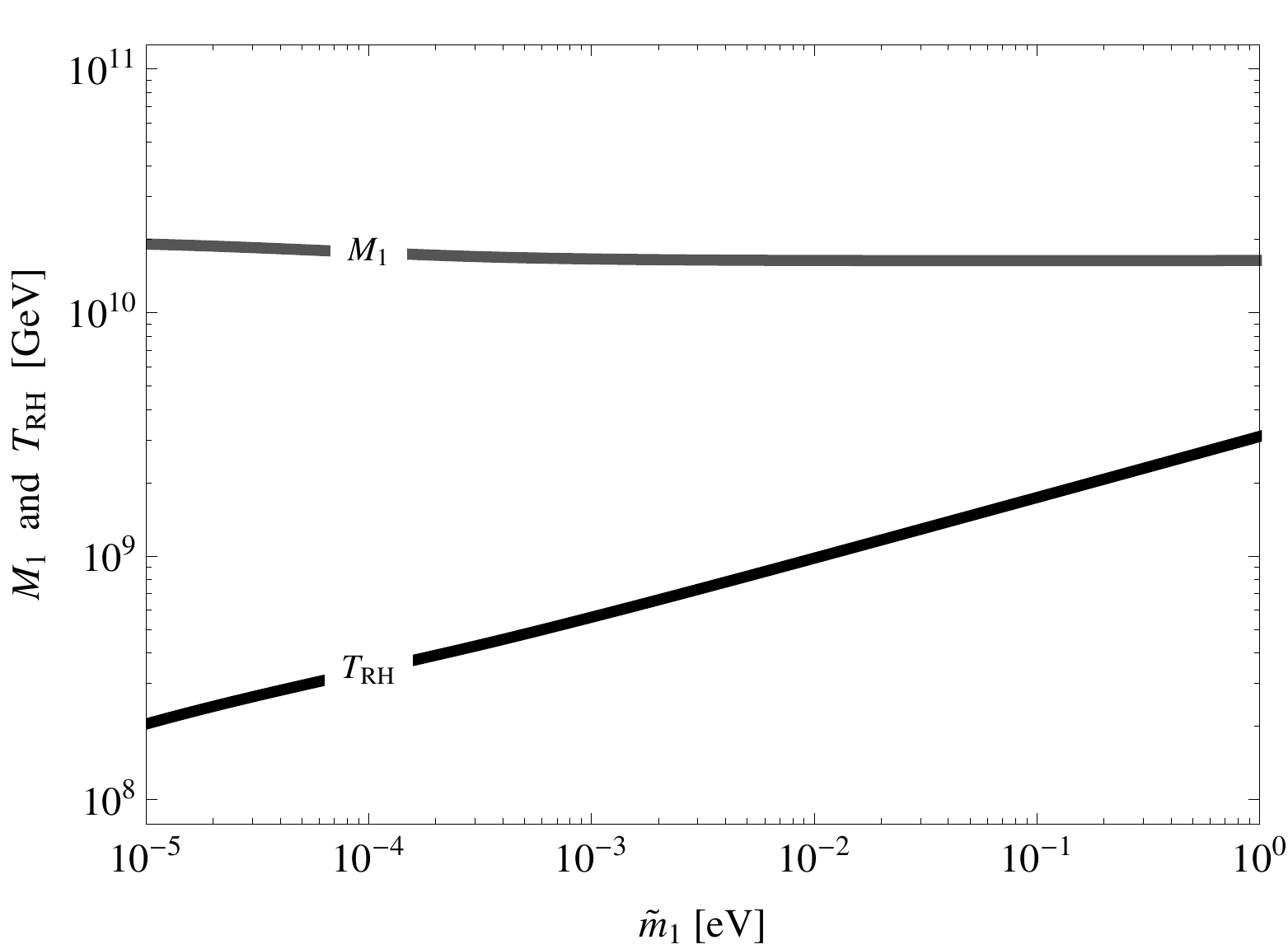}
\caption{Lower bounds on the heavy neutrino mass $M_1$ and the reheating
temperature $T_{\mathrm{RH}}$ as functions of the effective neutrino
mass $\widetilde{m}_1$ from successful leptogenesis.}
\label{fig:lbound}
\end{center}
\end{figure}

The reheating process is dominated by decays of the $B$$-$$L$ Higgs boson
$S$ into heavy neutrinos and the subsequent decay of these into Standard Model 
particles and their superpartners (cf.~Fig.~\ref{fig:reheat}, upper panel). 
As the detailed analysis of Ref.~\cite{Buchmuller:2012wn} shows, 
the competition between these decays and the cosmic expansion leads to an 
intermediate plateau of approximately constant `reheating temperature' 
$T_{\mathrm{RH}}(M_1,\widetilde{m}_1)$ (cf.~Fig.~\ref{fig:reheat}, 
lower panel), which is defined by $\Gamma^S_{N_1}(a_{\mathrm{RH}}) =
H(a_{\mathrm{RH}})$ where $H$ and $\Gamma^S_{N_1}$ are the Hubble parameter
and the effective decay rate of the $N_1$ neutrinos produced in $S$ decays, respectively. 
Note that this effective reheating temperature takes the dynamics of the reheating process into
account. Hence, it depends on the decay rates of $S$ and $N_1$,
and consequently on $M_1$ and $\widetilde{m}_1$, contrary to the 
mere decay temperature of the Higgs boson $S$, which would only depend
on $M_1$. Using $T_{\text{RH}}(M_1,\widetilde{m}_1)$ as a measure for the temperature
scale, the standard formula for thermal gravitino production is a good approximation. 
Successful leptogenesis implies lower bounds on $M_1$ and 
$T_{\text{RH}}(M_1,\widetilde{m}_1)$, which can be obtained by solving the
relevant set of Boltzmann equations. The results of the analysis in 
Ref.~\cite{Buchmuller:2012wn} are shown in Fig.~\ref{fig:lbound}.

\subsubsection*{LSP production from the thermal bath and in heavy gravitino decays}

The WIMP dark matter abundance from thermal freeze-out strongly depends on 
the nature of the LSP.
The mass spectrum of superparticles, motivated by anomaly mediation and
the present hints for the Higgs boson mass from LHC, has a characteristic
hierarchy \cite{Ibe:2011aa,Jeong:2011sg,Krippendorf:2012ir},
\begin{equation}\label{masshierarchy}
m_{\mathrm{LSP}}  \ll 
m_{\mathrm{squark},\mathrm{slepton}} 
\ll m_{\widetilde{G}} \, ,
\end{equation}
where $m_{\widetilde G}$ denotes the gravitino ($\tilde G$) mass.
Due to this hierarchy the LSP is typically a `pure' gaugino or higgsino.
It is well known that in this situation the thermal abundance of a bino LSP is
generically too large, which is therefore disfavoured. Hence, the case 
of a light wino \cite{Ibe:2011aa} or higgsino \cite{Jeong:2011sg} is 
preferred.\footnote{Note that a `pure' higgsino also occurs as next-to-lightest superparticle
alongside multi-TeV coloured particles in hybrid 
gauge-gravity mediation, however with the gravitino as LSP \cite{Brummer:2011yd}.\vspace{4pt}}
A pure neutral wino or higgsino is almost mass degenerate with a chargino
belonging to the same $\mathrm{SU(2)}$ multiplet. Hence, the current lower
bound on chargino masses \cite{Nakamura:2010zzi} also applies to the LSP.
The thermal abundance of a pure wino ($\widetilde w$) or higgsino ($\widetilde h$) LSP becomes only significant for 
masses above $1~\mathrm{TeV}$ where it is well approximated
by \cite{ArkaniHamed:2006mb}
\begin{equation}\label{dmth}
\Omega^{\mathrm{th}}_{\widetilde{w},\widetilde{h}} h^2 
= c_{\widetilde{w},\widetilde{h}} 
\left(\frac{m_{\widetilde{w},\widetilde{h}}}{1~\mathrm{TeV}}\right)^2 \ , 
\quad c_{\widetilde{w}} = 0.014\ , \quad c_{\widetilde{h}} = 0.10\ ,
\end{equation}
for wino\footnote{Compared to Ref.~\cite{ArkaniHamed:2006mb} we have reduced 
the abundance by 30\% to account for the Sommerfeld enhancement effect
\cite{Hisano:2006nn,Cirelli:2007xd}.} and higgsino, respectively.

Let us now consider gravitino masses in the range from 
$10~\mathrm{TeV}$ to $10^3~\mathrm{TeV}$, as suggested by anomaly mediation.
The gravitino lifetime is given by
\begin{equation}\label{Glife}
\tau_{\widetilde{G}} = \Gamma_{\widetilde{G}}^{-1} =
\left(\frac{1}{32\pi}\left(n_v + \frac{n_m}{12}\right)
\frac{m_{\widetilde{G}}^3}{M_{\mathrm{P}}^2}\right)^{-1}
= 24 \left(\frac{10~\mathrm{TeV}}{m_{\widetilde{G}}}\right)^3 \mathrm{sec}\ ,
\end{equation}
where $M_{\mathrm{P}} = 2.4\times 10^{18}~\mathrm{GeV}$, and $n_v = 12$ and
$n_m = 49$ are the number of vector and chiral matter multiplets, respectively.
The lifetime (\ref{Glife}) corresponds to the decay temperature
\begin{equation}
T_{\widetilde{G}} = \left(\frac{90~\Gamma_{\tilde G}^2~M_{\mathrm{P}}^2}{\pi^2 
g_*(T_{\widetilde{G}})}\right)^{1/4}
= 0.24 \left(\frac{10.75}{g_*(T_{\widetilde{G}})}\right)^{1/4}
\left(\frac{m_{\widetilde{G}}}{10~\mathrm{TeV}}\right)^{3/2} \mathrm{MeV}\,,
\end{equation}
with $g_*(T_{\tilde G})= 43/4$ counting the effective number of relativistic degrees of freedom.
For gravitino masses between $10~\mathrm{TeV}$ to $10^3~\mathrm{TeV}$
the decay temperature $T_{\widetilde{G}}$ varies between 
$0.2~\mathrm{MeV}$ and $200~\mathrm{MeV}$, i.e.\ roughly between the
temperatures of nucleosynthesis and the QCD phase transition. In this
temperature range the entropy increase due to gravitino decays and hence the corresponding dilution of the baryon asymmetry are negligible.

The decay of a heavy gravitino, $m_{\widetilde{G}} \gg m_{\mathrm{LSP}}$,
produces approximately one LSP. This yields the nonthermal contribution to the dark matter
abundance\footnote{Note that the thermal gravitino production rate has
a theoretical uncertainty of at least a factor of 2. The numerical prefactor used in Eq.~\eqref{dmG} was obtained by solving the Boltzmann equations governing the reheating process for $T_{\text{RH}} \in [ 10^8, 10^{11} ]$~GeV, cf.\ Ref.~\cite{Buchmuller:2012wn}. For an analytical approximation, see Appendix D in Ref.~\cite{Buchmuller:2010yy}.},
\begin{equation}\label{dmG}
\Omega_{\mathrm{LSP}}^{\widetilde{G}} h^2  =  \frac{m_{\mathrm{LSP}}}{m_{\widetilde{G}}}
\Omega_{\widetilde{G}} h^2 
 \simeq 2.7\times 10^{-2} 
\left(\frac{m_{\mathrm{LSP}}}{100~\mathrm{GeV}}\right)
\left(\frac{T_{\mathrm{RH}}(M_1,\widetilde{m}_1)}{10^{10}~\mathrm{GeV}}
\right)\ ,
\end{equation}
where we have assumed that the gravitino density is produced from the
thermal bath during reheating, cf.\ Fig.~\ref{fig:reheat}, upper panel. 
For LSP masses below $1~\mathrm{TeV}$, which are most interesting for
the LHC as well as for direct searches, the total LSP abundance
\begin{equation}\label{dmtot}
\Omega_{\widetilde{w},\widetilde{h}} h^2 =
\Omega_{\widetilde{w},\widetilde{h}}^{\widetilde{G}} h^2 + 
\Omega_{\widetilde{w},\widetilde{h}}^{\mathrm{th}} h^2
\end{equation} 
is thus dominated by the contribution from gravitino decay.

The LSPs are produced relativistically. They form warm dark matter which can affect structure formation on small scales. A straightforward calculation yields the free-streaming length
\begin{equation}
 \lambda_{FS} = \int_{\tau_{\tilde G}}^{t_0} dt \frac{v_{\text{LSP}}}{a} \simeq \left(\frac{3}{4} \right)^{2/3}
 \frac{m_{\tilde G}}{2 \, m_{\text{LSP}}} \left(\tau_{\tilde G} \, t_{eq} \right)^{1/2} \left(\frac{t_0}{t_{eq}} \right)^{2/3} \left( \ln \frac{16 \, t_{eq} \, m^2_{\text{LSP}}}{\tau_{\tilde G} \, m^2_{\tilde G}} + 4 \right) \,,
\end{equation}
where $t_{eq}$ and $t_0$ denote the time of radiation-matter-equality and the age of the universe, respectively. For the gravitino and LSP masses considered in this paper, one finds $\lambda_{FS} \lesssim 0.1$~Mpc, which is below the scales relevant for structure formation \cite{Borzumati:2008zz}.

\begin{figure}[t]
 \begin{center}
  \includegraphics[width=12cm]{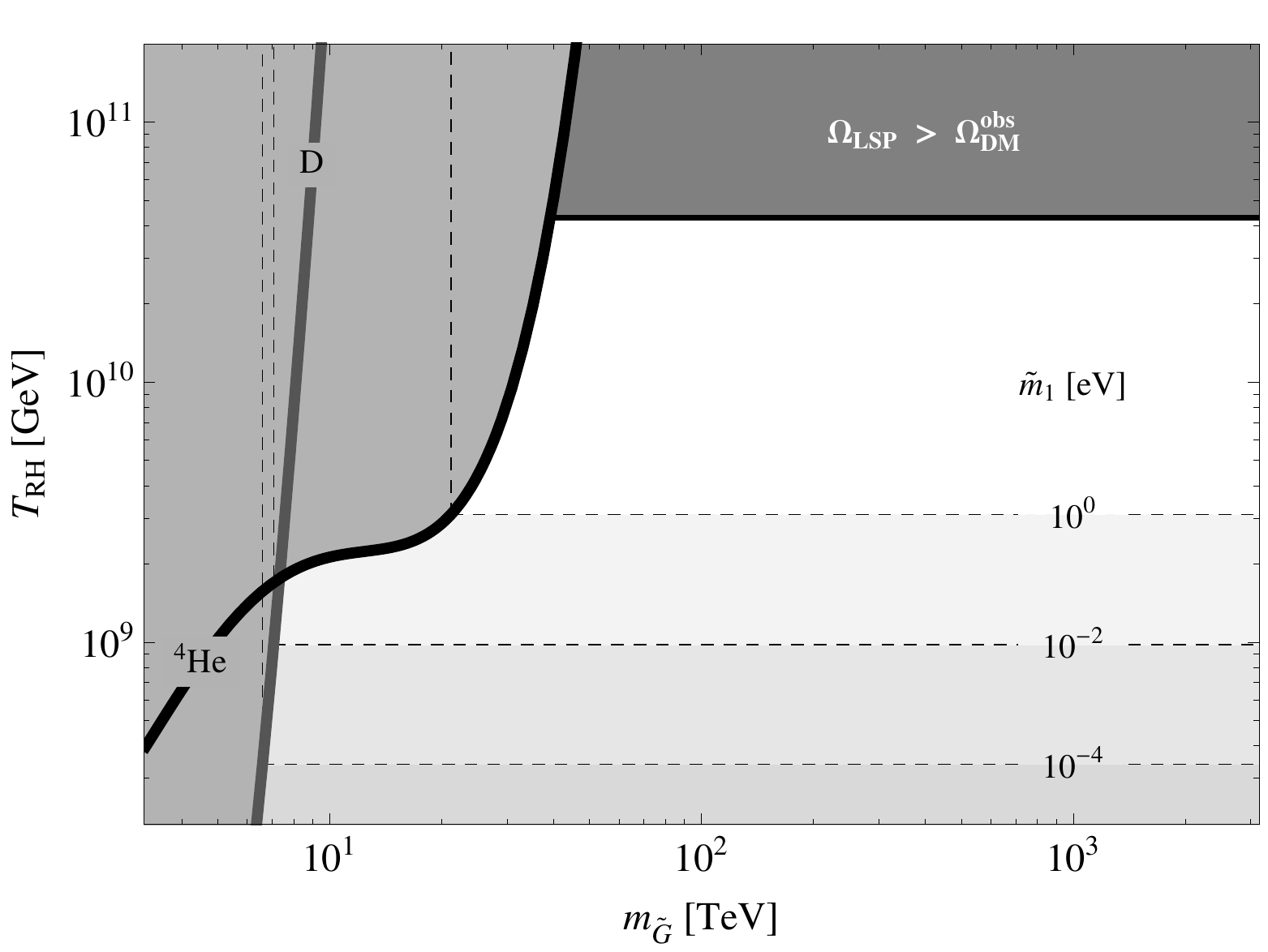}\\
\caption{Upper and lower bounds on the reheating
temperature as functions of the gravitino mass.
The horizontal dashed lines denote lower bounds imposed by successful leptogenesis for different values of the effective neutrino mass $\widetilde m_1$,
cf.~Fig.~\ref{fig:lbound} and Ref.~\cite{Buchmuller:2012wn}. The curves labelled $^4\text{He}$
and D denote upper bounds originating from the primordial helium-4 and deuterium abundances created during BBN, which are taken from \cite{Kawasaki:2008qe} (case~2, which
gives the most conservative bounds).
The vertical dashed lines represent the absolute lower bounds on the gravitino mass for fixed
effective neutrino mass $\widetilde{m}_1$ and minimal reheating temperature.
The shaded region marked $\Omega_{\text{LSP}} > \Omega^{\text{obs}}_{\text{DM}}$ is excluded as it corresponds to overproduction of dark matter, taking into account that the LSP mass is bounded from below, $m_{\mathrm{LSP}} \geq 94~\mathrm{GeV}$
(see text).}
\label{fig:Thwbounds1}
 \end{center}
\end{figure}

\subsubsection*{Relations between LSP, gravitino and neutrino masses}

The LSP has to be heavier than $94~\mathrm{GeV}$, the current lower bound 
on chargino masses \cite{Nakamura:2010zzi}. From the requirement of LSP
dark matter, i.e.\
$\Omega_{\mathrm{LSP}} h^2 = \Omega_{\mathrm{DM}} h^2 \simeq 0.11$
\cite{Nakamura:2010zzi}, one then obtains an upper bound on the reheating 
temperature, $T_{\mathrm{RH}} < 4.2\times 10^{10}~\mathrm{GeV}$.
For gravitino masses below $40~\mathrm{TeV}$, primordial nucleosynthesis
provides a more stringent upper bound on the reheating temperature 
\cite{Kawasaki:2008qe}. In Fig.~\ref{fig:Thwbounds1} we compare 
upper and lower
bounds on the reheating temperature from dark matter density, nucleosynthesis
and leptogenesis, respectively, as functions of the gravitino mass. It is
remarkable that for the entire mass range, 
$10~\mathrm{TeV} \lesssim m_{\widetilde{G}} \lesssim 10^3~\mathrm{TeV}$,
nucleosynthesis, dark matter and leptogenesis can be consistent.

\begin{figure}[t]
\subfigure{
 \includegraphics[width=0.48\textwidth]{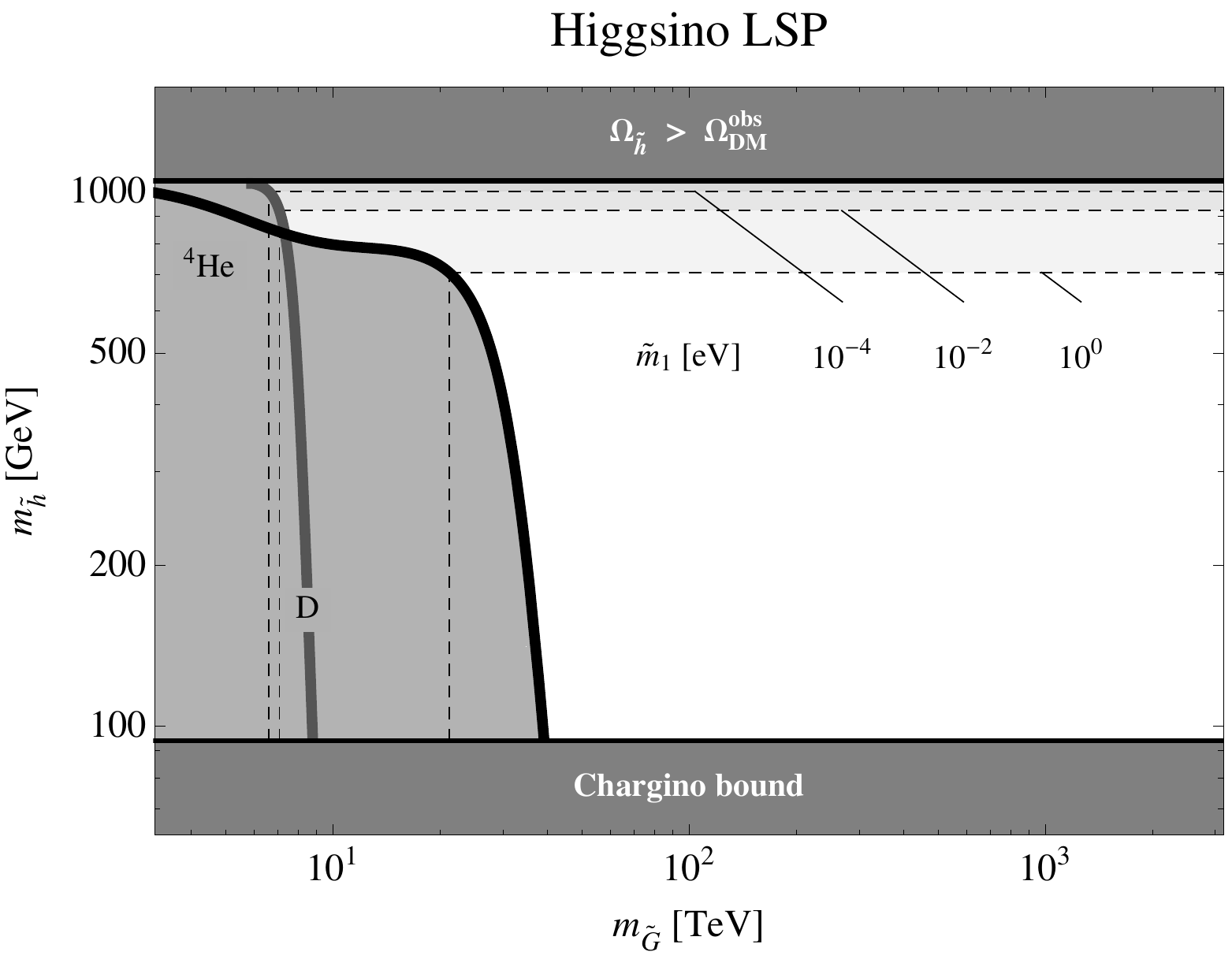}}
\subfigure{
 \includegraphics[width=0.48\textwidth]{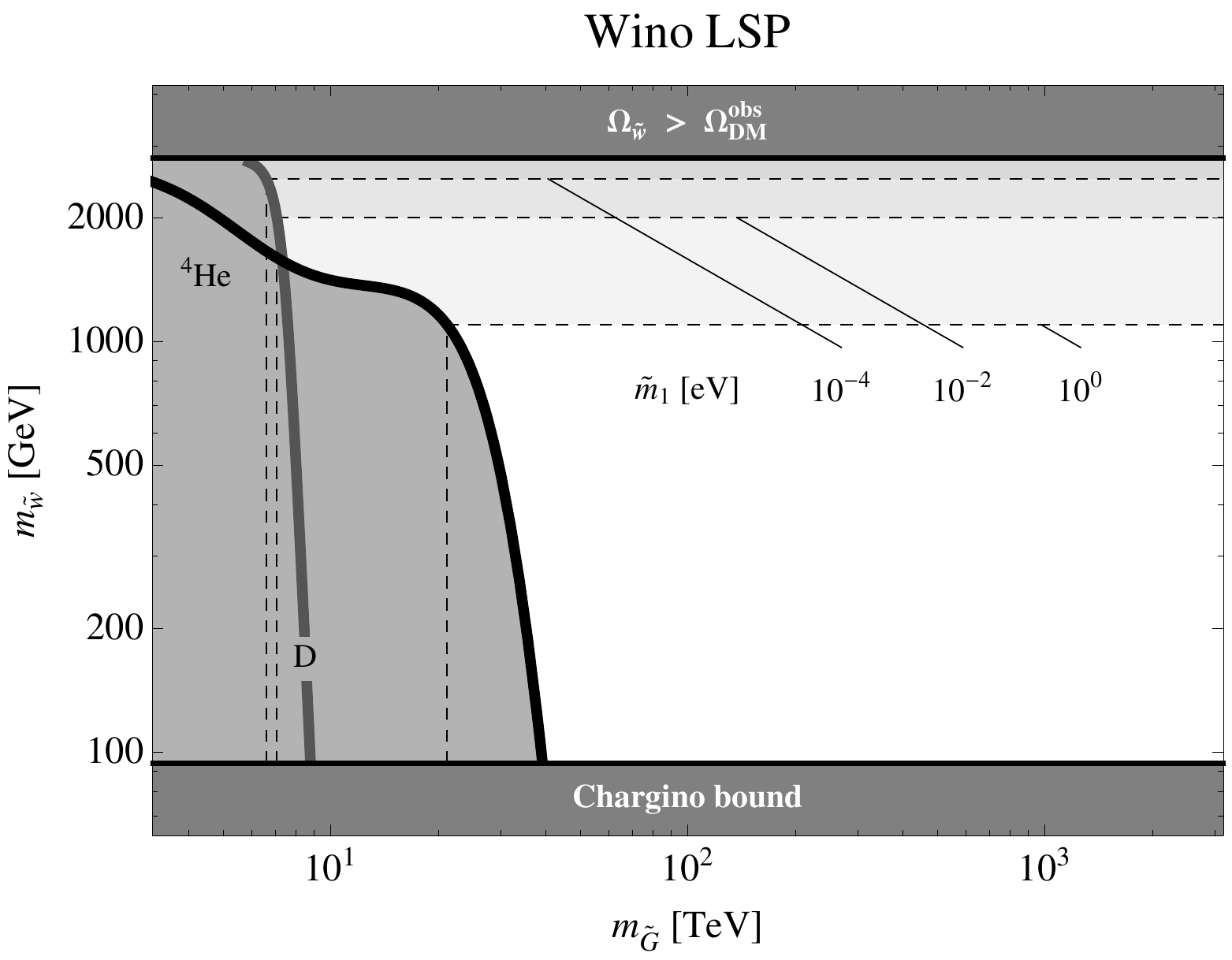}}
\caption{Upper and lower bounds on the LSP mass in the higgsino and wino
case, respectively, and lower bounds on the gravitino mass.
These bounds are in one-to-one correspondence with the bounds on the
reheating temperature and the gravitino mass in Fig.~\ref{fig:Thwbounds1}.
The horizontal dashed lines denote the upper bounds on the LSP mass imposed by successful leptogenesis for different values of the effective neutrino mass $\widetilde m_1$.
The curves labelled $^4\text{He}$ and D denote lower bounds on the LSP as well as on the
gravitino mass originating from the primordial helium-4 and deuterium abundances created during BBN.
The vertical dashed lines represent the absolute lower bounds on the gravitino mass for fixed
effective neutrino mass $\widetilde{m}_1$ and maximal LSP mass.
The dark shaded regions on the upper edge of the plots correspond to thermal overproduction of dark matter and are hence excluded.
We do not consider LSP masses below $94\,\textrm{GeV}$ due to the present lower bound on the
chargino mass (see text).
}
\label{fig:Thwbounds2}
\end{figure}

The dark matter constraint $\Omega_{\mathrm{LSP}} h^2 = \Omega_{\mathrm{DM}} h^2 \simeq 0.11$,
with $\Omega_{\mathrm{LSP}} h^2$ calculated according to Eqs.~(\ref{dmth}), (\ref{dmG}) and Eq.~(\ref{dmtot}), establishes a one-to-one connection between LSP masses and values
of the reheating temperature.
This relation maps the viable region in the
$\left(m_{\widetilde{G}},T_{\textrm{RH}}\right)$-plane for a given effective neutrino mass $\widetilde{m}_1$ into the corresponding viable region in the
$\left(m_{\widetilde{G}},m_{\textrm{LSP}}\right)$-plane.
We present our results for higgsino and wino LSP in the two panels of
Fig.~\ref{fig:Thwbounds2}, respectively.
The upper bound on the LSP mass is a consequence of the lower bound on the reheating
temperature from leptogenesis, which is why it depends on the effective neutrino mass $\widetilde{m}_1$.
The lower bound on the LSP mass corresponds to the upper bound on the reheating temperature
from BBN and hence depends on the gravitino mass $m_{\widetilde{G}}$.
This latter relation between $m_{\textrm{LSP}}$ and $m_{\widetilde{G}}$ can also be
interpreted the other way around.
As each LSP mass is associated with a certain reheating temperature, we find for
each value of $m_{\textrm{LSP}}$ a lower bound on the gravitino mass.
For given $\widetilde{m}_1$ we then obtain an absolute lower bound on the gravitino mass
by raising the LSP mass to its maximal possible value.

\begin{figure}
\begin{center}
\includegraphics[width=12cm]{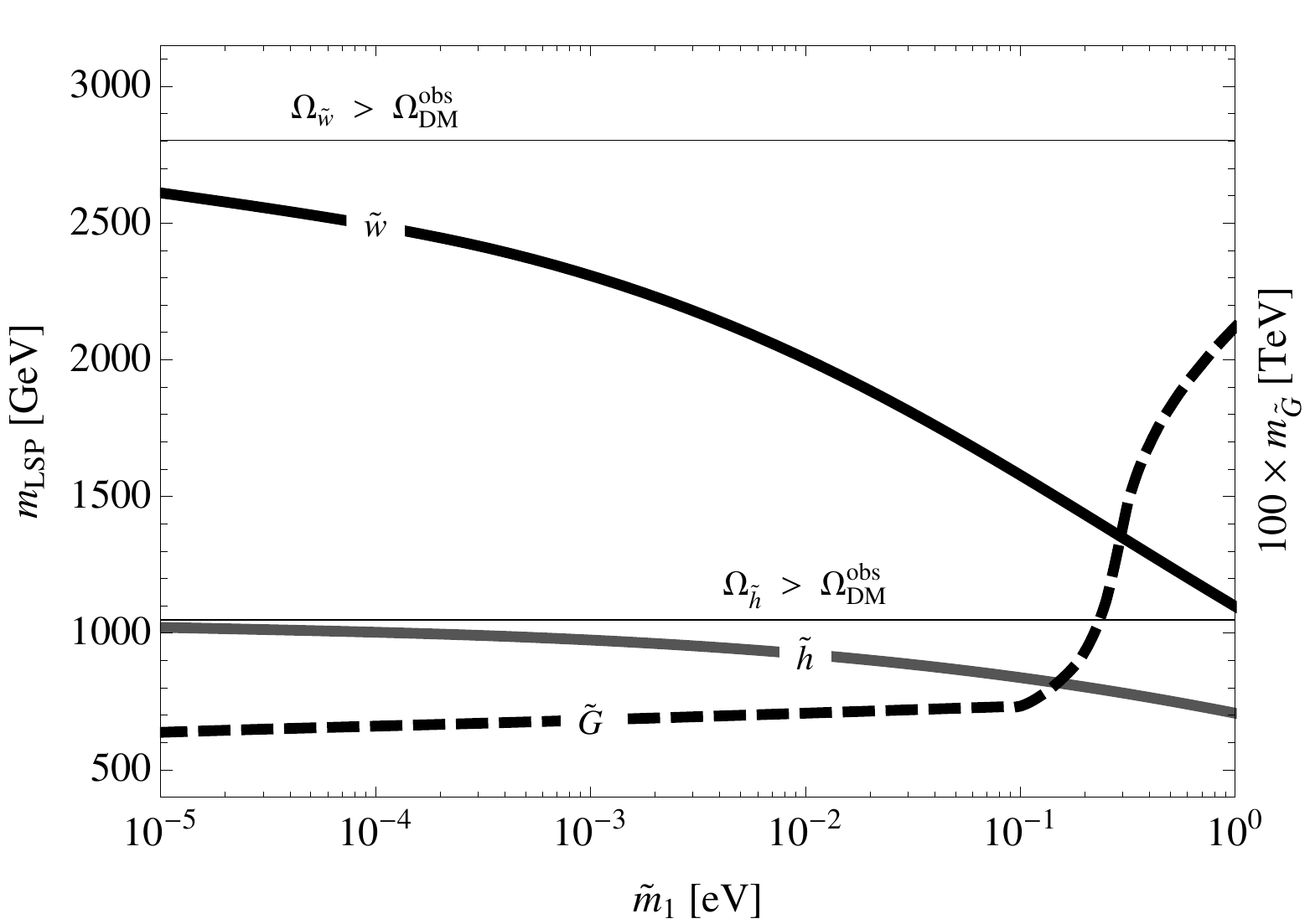}
\caption{Upper bounds on wino ($\widetilde{w}$) and higgsino
($\widetilde{h}$) LSP masses imposed by successful leptogenesis as well as absolute
lower bound on the gravitino mass according to BBN
as functions of the effective neutrino mass $\widetilde{m}_1$.
Note that in Fig.~\ref{fig:Thwbounds2} these bounds are indicated by horizontal
and vertical dashed lines, respectively, for different value for $\widetilde{m}_1$.
Wino masses larger than $2.8\,\textrm{TeV}$ and higgsino masses larger than
$1.0\,\textrm{TeV}$ result in thermal overproduction.}
\label{fig:lbound2}
\end{center}
\end{figure}

The upper bound on the LSP mass as well as the absolute lower bound on the gravitino
mass both depend on the effective neutrino mass $\widetilde{m}_1$.
In Fig.~\ref{fig:lbound2} we now finally show the explicit dependence of
these bounds on $\widetilde{m}_1$.
The upper bound on the LSP mass imposed by successful leptogenesis increases when lowering
$\widetilde{m}_1$, i.e.\ when extending the range of allowed reheating temperatures to lower
values.
For very small $\widetilde{m}_1$ it approaches the upper bound on the LSP mass above which
thermal freeze-out leads to an overabundance of LSPs.
At large values of $\widetilde{m}_1$, the bound on the LSP mass from leptogenesis
becomes stronger.
Furthermore, we find that the absolute lower bound on the gravitino mass is
rather insensitive to the effective neutrino mass for $\widetilde m_1 \lesssim 10^{-1}$~eV, but rapidly increases as a function of $\widetilde{m}_1$ for larger values of $\widetilde m_1$.
This reflects the fact that small values of $\widetilde m_1$ correspond to low
reheating temperatures, for which the allowed range of gravitino masses,
being determined by the BBB abundance of deuterium, hardly changes with when varying the
temperature.
It turn, when the allowed range of gravitino masses is determined by the BBN abundance
of helium-4, which is the case for very large $\widetilde{m}_1$, the absolute lower bound on
$m_{\widetilde{G}}$ increases with $\widetilde{m}_1$.

\subsubsection*{Prospects for direct detection and collider experiments}

For pure wino and higgsino LSPs, the exchange of the lightest Higgs boson
yields at tree level for the spin-independent elastic scattering cross section 
\cite{Hisano:2004pv},
\begin{align}
\sigma^{\widetilde{w}}_{\mathrm{SI}} &\sim 2\times 10^{-43}~\mathrm{cm^2}
\left(\frac{125~\mathrm{GeV}}{m_{h^0}}\right)^4 
\left(\frac{100~\mathrm{GeV}}{m_{\widetilde{h}}}\right)^2
\left(\sin{2\beta}+\frac{m_{\widetilde{w}}}{m_{\widetilde{h}}}\right)^2\ ,\\
\sigma^{\widetilde{h}}_{\mathrm{SI}} &\sim 7\times 10^{-44}~\mathrm{cm^2}
\left(\frac{125~\mathrm{GeV}}{m_{h^0}}\right)^4 
\left(\frac{100~\mathrm{GeV}}{m_{\widetilde{w}}}\right)^2\ ,
\end{align}
where $m_{h^0}$ is the mass of the lightest Higgs boson. For the hierarchical
mass spectrum of Eq.~(\ref{masshierarchy}) one has 
$r_{\widetilde{w}} \equiv m_{\widetilde{w}}/m_{\widetilde{h}} \ll 1$ for
wino LSP and
$r_{\widetilde{h}} \equiv m_{\widetilde{h}}/m_{\widetilde{w}} \ll 1$ for higgsino LSP, respectively. Hence, the spin-independent
scattering cross sections are significantly below the present experimental
sensitivity for LSP masses below $1~\mathrm{TeV}$.

For the considered hierarchy of superparticle masses, gluinos and squarks
are heavy. Hence the characteristic missing energy
signature of events with LSPs in the final state may be absent and the discovery
of winos or higgsinos therefore very challenging \cite{lhc}. In both cases
the neutral LSP is almost mass degenerate with a chargino, which increases
the discovery potential. One may hope for  macroscopic charged tracks
of the produced charginos. A generic prediction is also the occurrence of
monojets caused by the Drell-Yan production of higgsino/wino pairs 
associated by initial state gluon radiation.

\subsubsection*{Conclusion}

We have shown that spontaneous breaking of $B$$-$$L$ symmetry can successfully generate the initial conditions for the hot early universe, i.e.\ entropy, baryon asymmetry and dark matter, for the hierarchical superparticle mass spectrum given in Eq.~\eqref{masshierarchy}. Very heavy gravitinos, as motivated by hints for the Higgs boson at the LHC, are produced from the thermal bath during the reheating phase after inflation. They eventually decay at some time between the QCD phase transition and BBN into wino or higgsino LSPs, which then account for the observed dark matter abundance. By additionally imposing the requirement of successful leptogenesis, we obtain upper bounds on the LSP masses and a lower bound for the gravitino mass. We emphasize that the initial conditions of the radiation dominated phase of the early universe, in particular the reheating temperature, are not free parameters but are determined by parameters of a Lagrangian, which in principle can be measured by particle physics experiments and astrophysical observations.

\vspace{0.5cm}
\subsubsection*{Acknowledgements}

The authors thank T.~Bringmann and F.~Br\"ummer for helpful discussions and comments. This
work has been supported by the German Science Foundation (DFG) within the Collaborative
Research Center 676 ``Particles, Strings and the Early Universe''.

\end{document}